# Developing Smart Web-Search Using RegEx

Ikechukwu Onyenwe[1], Stanley Ogbonna[1], Ebele Onyedimma[1], Onyedikachukwu Ikechukwu-Onyenwe[1], Chidinma Nwafor[2]

[1]Nnamdi Azikiwe University, Awka, Nigeria
[2]Nigerian Army College Of Environmental Science And Technology, Makurdi, Benue State
Corresponding Author: ie.onyenwe@unizik.edu.ng

**ABSTRACT**: Due to the increasing storage data on the Web Applications it becomes very difficult to use only keyword-based searches to provide comprehensive search results, thus increasing the difficulty for web users to search information on the web. In this paper, we proposed using a combined method of keyword-based and Regular expressions (regEx) searches to perform search using strings of targeted items for optimal results even as the volume of data around the world on the Internet continues to explode. The idea is to embed regEx patterns as part of search engine's algorithm in a web application project to provide strings related to the targeted items for more comprehensive coverage of search results. The user's search query is a string of characters guided by search boundaries selected from the entry point. The results returned from the search operation are different results within a category determined by the search boundaries. This is designed to be beneficial to a user who has an obscure idea about the information he/she wanted to search but knows the boundries within which to get the information. This technique can be applied to data processing tasks such as information extraction and search refinement.

## I. INTRODUCTION

Web-search involves searching for information on the Web. The search results are generally presented in a line of results often referred to as search engine results pages (SEROs). The information may be a mix of web pages, images, and other types of files. An entry point of most web-search is a search box/bar.

A search box is a graphical control element used in applications such as file managers, web browsers and websites. A search box is usually a single-line textbox with the dedicated function of accepting user input to be searched for in a database. On web pages users are allowed to enter a query to be submitted to a web-search engine server-side script, where an index database is queried for entries that contain one or more of the user's keyword research. it is an integral part of the site search functionality, which is an important element of website design for content-rich websites. On some websites, site search is more prominent than on others. E-commerce websites typically use search boxes, and thus site search, as a primary navigation tool.

Regular expression (regex) is based on the concept of a state machine, which is a process that will sequentially read in the symbols of an input word and decide whether the current state of the machine is one of acceptance or non-acceptance [3]. Using a regex requires defining all of the criteria that need to reach an accepted state in order for a valid pattern match to occur. Illustrating regex concept, we used example from [3] where *XYZ* is a regex to perform a pattern match on the string *ZXYXYZ*. The regex solution starts by reading *Z* symbol into the state machine which failed to match the first condition of the regex pattern thus causing a state of non-acceptance and hence failure of the matching operation. The next characters X and Y in the string matched the first and second condition of the regex but failed at the character *Y* to reach an accepted state. This string portion starting with second character *X* failed to reach an accepted state. The third character of the string *Y* is used as a start symbol for the state machine which failed to match the first condition of the start machine and results in a failure as well. Finally, the last three characters X, Y and X of the string read into the state machine successfully match the XYZ regex conditions.

Getting a comprehensive search result on the Web even when the user is not sure of the right search keywords to use is an important module required for developing a user-friendly web application. The explosive rate of information growth and availability often makes it increasingly difficult to locate information pertinent to users needs. Use of only keyword based search methodologies are not adequate for describing the information users seek. In this paper, we propose the use of a search method that combined keyword-based and regex searches for an efficient but comprehensive search output. We used keywords to serve as search boundaries and strings to match and filter all items pertinent to what the users seek within a confined boundary. This technique can be applied to data processing tasks such as information extraction and search refinement.

According to [**2**], The Web search engines started to take form with the use of regular expressions to search through their indexes. Regular expressions were chosen for these early search engines because of both their power and easy implementation. It is a fairly trivial task to convert search strings into regular expressions that accept only strings that have some relevance to the query. In the case of a search engine, the strings input to the regular expression would be either whole web pages or a pre-computed index of a web page that holds only the most important information from that web page. A query such as regular expression in Fig. 1 could be translated into the following regular expression, then, of course, would be the set of all characters in the character encoding used with this search engine. The results returned to the user would be the set of web pages that were accepted by this regular expression. [**4**] focused on text preprocessing of automotive advertisements domains to configure a structured database. The structured database was created by extract the information over unstructured automotive advertisements, which is an area of natural language processing. Information extraction deals with finding factual information in text using learning regular expressions. [**5**] presents the automation of a Web advertising recognition algorithm, using regular expressions. The tests were carried out in three Web browsers. As a result, the detection of advertisements in Spanish, that distract attention and that above all extract information from users was achieved. Extraction of information from the Web is a well known but unsolved and critical problem when it comes to accessing complex information systems. These problems are related to the extraction, management and reuse of the huge amount ofWeb data available. These data usually has a high heterogeneity, volatility and low quality (i.e. format and content mistakes), so it is quite hard to build reliable systems. [**1**] proposed an Evolutionary Computation approach to the problem of automatically learn software entities based on Genetic Algorithms and regular expressions. One use of regular expressions that used to be very common was in web search engines. Archie, one of the first search engines, used regular expressions exclusively to search through a database of filenames on public FTP servers. [**6**] presented a novel approach for generating string test data for string validation routines, by harnessing the Internet. The technique uses program identifiers to construct web search queries for regular expressions that validate the format of a string type (such as an email address). It then performs further web searches for strings that match the regular expressions, producing examples of test cases that are both valid and realistic. Following this, our technique mutates the regular expressions to drive the search for invalid strings, and the production of test inputs that should be rejected by the validation routine.

$$(\Sigma^*\text{regular}\Sigma^*\text{expression}\Sigma^*)^* \cup (\Sigma^*\text{expression}\Sigma^*\text{regular}\Sigma^*)^*$$

Fig. 1 : Regex query. Source [**Carter**]

## 2. Experimental Tools

The followings describe the tools we used in this paper. Basically, we used Django, Python and Regular Expression commonly known as regex. We have introduced and extensively discuss regex in the introduction section.

**Django** is a Python-based free and open-source web framework that follows the model–template–views (MTV) architectural pattern. It is maintained by the Django Software Foundation (DSF), an American independent organization established as a non-profit. Django's primary goal is to ease the creation of complex, database-driven websites. The framework emphasizes reusability and "pluggability" of components, less code, low coupling, rapid development, and the principle of do not repeat yourself (DRY). Python is used throughout, even for settings, files, and data models. Django also provides an optional administrative create, read, update and delete interface that is generated dynamically through introspection and configured via admin models.

**Python** is an interpreted high-level general-purpose programming language. Its design philosophy emphasizes code readability with its use of significant indentation. Its language constructs as well as its object-oriented approach aim to help programmers write clear, logical code for small and large-scale projects. Python is dynamically-typed and garbage-collected. It supports multiple programming paradigms, including structured (particularly, procedural), object-oriented and functional programming. It is often described as a "batteries included" language due to its comprehensive standard library.

## 3. Results and Discussion

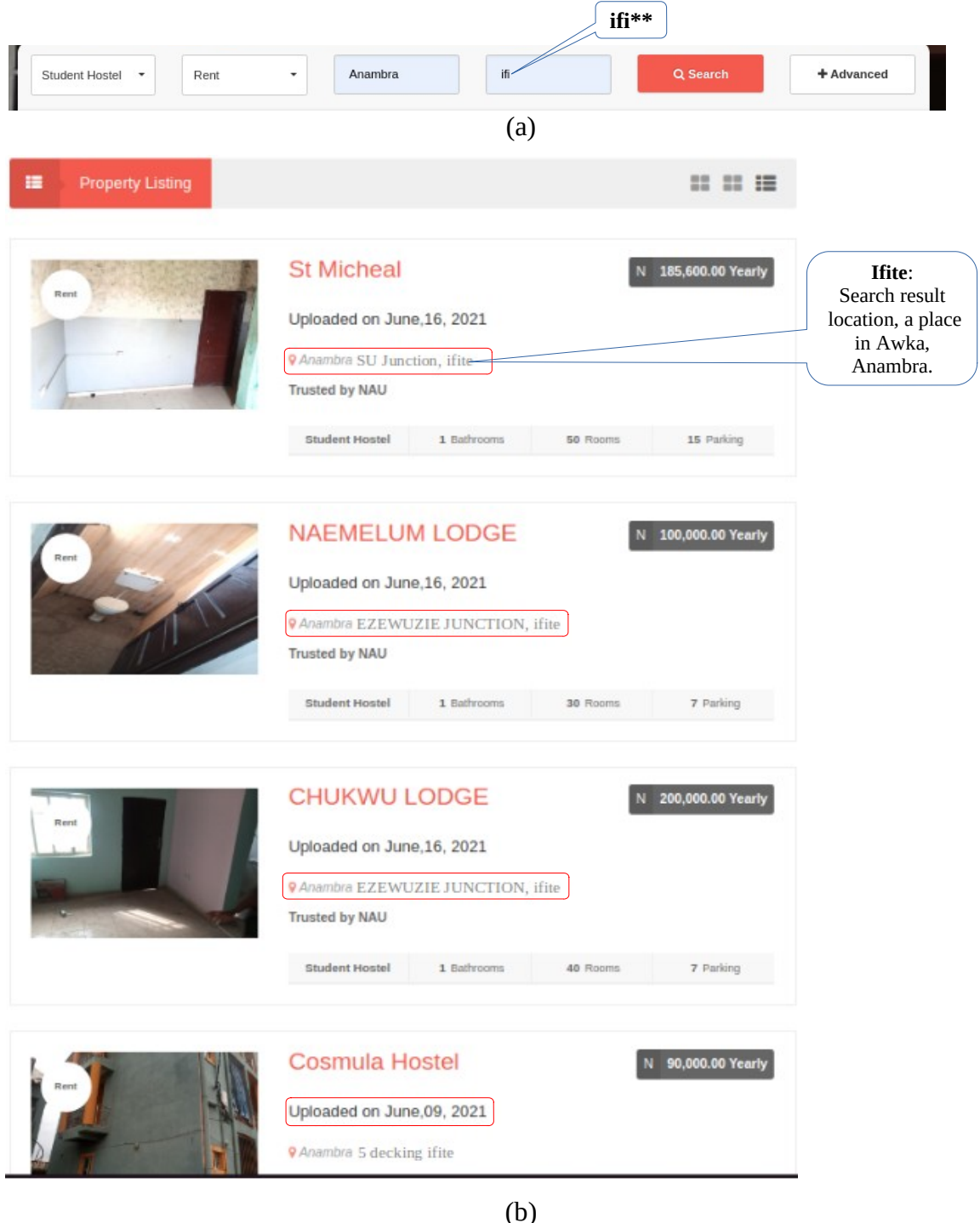

Fig. 2: Search box and search results using propertywithin.com.ng.

For this work, we used a property network website known as **propertywithin.com.ng** and integrate our combo search engine component in it. P**ropertywithin** is a friendly accommodation hub (search engine) aimed to facilitate users ability to find the property most suited to their needs regardless of location and position swiftly. There are subscribed trusted agents that bring onboard their properties for easy identification by possible occupants and create a customer-friendly atmosphere to choose property without much stress of physical sighting across Nigeria (including accommodation needs around campuses). It is an African real estate management agency that facilitates ease of identifying properties for both leasing, sales (buying), management, and securing of property' lives.

Fig. 2 (a) is the search box of the propertywithin website. The search engine here combined the use of keyword-based and regex-based searches. From this Figure, observe the keywords **{Student Hostel, Rent, Anambra}** and string **{ifi}**. We used the **keywords** to set our search boundaries to be *only student hostels for rent within a location in Awka Anambra* and the **string** to be *a regex pattern to match and filter all locations that have available accommodations.* The search results of Fig. 2 (a) is displayed in Fig. 2 (b). Take note of the last words in the red marked boxes in Fig. 2 (b). The words indicate that the found accommodations are in Ifite, a place in Awka of Anambra most populated by the students of Nnamdi Azikiwe University. Also note that there are streets/closes we don't hve an idea about them but they are populated alonside their accommodations.

### 4. Conclusion

In this paper, we have shown how combination of the keyword-based and regular expression (regEx)-based searches can be used to improve search efficacy in a comprehensive way. Illustrations in Fig.1 and 2 have shown how regEx based pattern matching is a highly useful tool for augmenting keyword based search methodologies in that it can be used to dramatically reduce the number of irrelevant search results the user will be presented with. Again, the search results are comprehensive enough that even the locations the user has an obscure idea about the information he/she wanted to search are also populated.